\pdfoutput=1
\documentclass[conference]{IEEEtran}
\usepackage{fancyhdr}

\usepackage{url}
\usepackage{graphicx}

\title{GarageWorks: A Shared Testbench Abstraction for Software Simulation and FPGA Testing}

\author{
\IEEEauthorblockN{Kazutomo Yoshii}
\IEEEauthorblockA{
Mathematics and Computer Science Division\\
Argonne National Laboratory\\
Lemont, Illinois, USA
}
}

\begin{document}
\maketitle


\begin{abstract}
GarageWorks is a lightweight FPGA testing framework that enables the same cocotb testbench to be used for both software RTL simulation and FPGA-accelerated hardware simulation. A shared abstraction hides backend-specific interfaces, while synthesizable Advanced eXtensible Interface (AXI) dispatchers provide design-under-test (DUT)-specific control and test operations on the FPGA. GarageWorks is intended for rapid bare-metal testing of standalone designs and on-chip processing logic for scientific sensors, rather than full SoC-level prototyping. The framework has been validated on AMD Alveo V80 FPGAs with several internal designs and provides initial support for AI-assisted generation of DUT-specific test components.
\end{abstract}

\section{Introduction}

Scientific instruments increasingly rely on custom logic designs such as data compression and feature detection for on-chip or near-sensor processing to handle  large volumes of streaming sensor data~\cite{miceli2009application,apsupgrade, miceli2022pushing, hammer2020strategies, Strempfer:2021uo}. Similarly, domain-specific accelerator logic can provide substantial performance benefits over CPUs and GPUs for some HPC workloads, such as molecular dynamics~\cite{shaw2021anton, yang2019fully} and Monte Carlo cross-section lookup kernel~\cite{yoshii2023hardware, yoshii2025case}. We have been prototyping both types of designs in-house for feasibility studies and eventual deployment, and expect the number and diversity of such designs to continue increasing. Improving the efficiency of the hardware design and prototyping flow is therefore becoming increasingly important in our work.

For custom logic development at the RTL level, software RTL simulation remains essential. Conventional Verilog testbenches provide a standard approach for Verilog designs, while ChiselSim provides an effective environment for simulating modules written in Chisel~\cite{Bachrach:ft}. cocotb~\cite{rosser2018cocotb} is often attractive for integration testing because Python makes it easy to build software reference models, generate and manipulate test data, use existing numerical libraries, and analyze and report results. RTL simulation also provides excellent observability and debugging capability.

FPGA-accelerated execution can be orders of magnitude faster than software RTL simulation~\cite{karandikar2018firesim}, enabling longer-running and more realistic tests.
FPGAs provide a practical platform for this purpose because the DUT and supporting test structures can execute together in reconfigurable hardware. However, moving from an RTL testbench to FPGA execution often requires substantial additional bring-up effort.

We introduce \textbf{GarageWorks}, a shared testbench abstraction layer for software RTL simulation and FPGA testing. GarageWorks supports DUTs written in Chisel, which is our recommended flow, as well as Verilog. By providing a tested common abstraction, it also simplifies AI-assisted generation of DUT-specific testbench components. Our goal is to support rapid hardware-assisted unit and integration testing while preserving the ability to switch back to RTL simulation when debuggability is needed.

\section{GarageWorks Abstraction and Design}

GarageWorks consists of three main parts: a shared testbench abstraction for
software RTL simulation and FPGA-accelerated hardware simulation, a common
configuration and backend mechanism that enables the same cocotb testbench to
target both environments, and AI-assisted generation of DUT-specific test
components. The FPGA side uses a user-defined synthesizable AXI
dispatcher to
implement DUT-specific control, data movement, and test sequencing, while the
software side uses a common bridge interface to hide backend-specific access
mechanisms.
GarageWorks grew from the open-source \texttt{chisel-axi-utils} project, originally developed to provide lightweight AXI4 bus-functional models (BFM) and interface utilities for Chisel. It was later extended with cocotb support, FPGA bridge functionality, parameter sharing, reset mechanisms, and FPGA-oriented templates.

\subsection{Shared Testbench Abstraction}

\begin{figure}[t]
    \centering
    \includegraphics[width=\columnwidth]{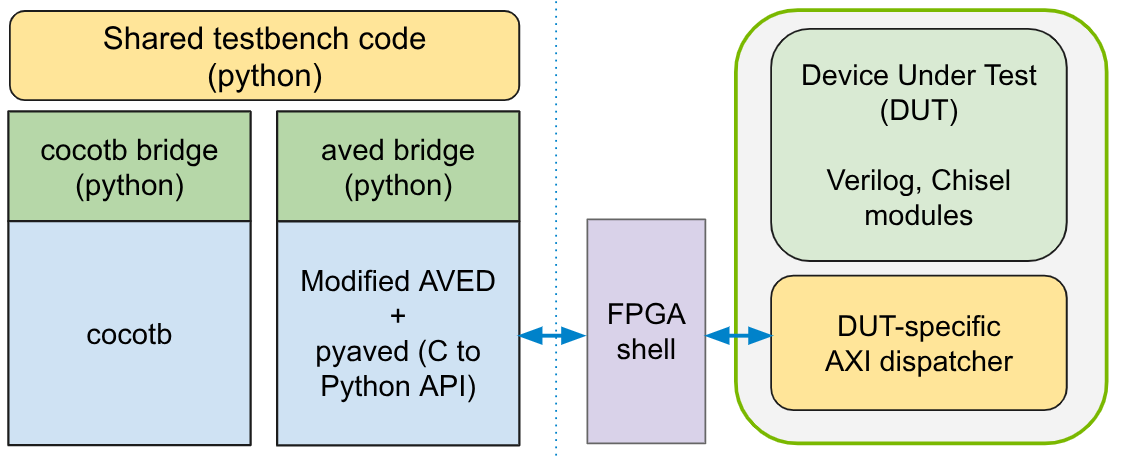}
    \caption{GarageWorks Architecture.}
    \label{fig:GarageWorks_arch}
\end{figure}

GarageWorks targets standalone domain-specific accelerators and
specialized on-chip processing blocks, such as data-processing logic for X-ray
detector ASICs, rather than full-featured SoC-level prototyping environments 
targeted by frameworks such as Chipyard~\cite{amid2020chipyard} and FireSim~\cite{karandikar2018firesim}.
Its lightweight design requires only DUT-specific test infrastructure, leaving
more FPGA resources available for the DUT and test-specific hardware
structures. Figure~\ref{fig:GarageWorks_arch} illustrates the overall CPU--FPGA
architecture and the relationship between the host-side software stack,
FPGA-side test infrastructure, and DUT.

Unlike regular testbenches, GarageWorks testbenches do not directly manipulate
DUT clocks or internal registers. To enable this, each DUT is connected to a
user-defined, synthesizable \emph{AXI dispatcher}, which receives host commands,
translates them into DUT-specific control and data operations, and is exposed
to the host as memory-mapped I/O. GarageWorks provides reusable synthesizable
AXI dispatcher modules that users can compose and customize for each DUT.

The AXI dispatcher accommodates DUT-specific test structures such as control
and status registers, input and output queues, memories, and start/stop
controls. 
The dispatcher also serves as a hardware-oriented test sequencer: depending on the
DUT and test case, sequencing can be implemented with hardcoded finite-state
machines, input/output queues, register or memory accesses, a lightweight
software-controlled CPU core, or a combination of these mechanisms. For
example, queue-based sequencing is well suited to image-processing tests, where
test images and processed outputs can be buffered directly on the FPGA. For a
CPU-like DUT, AXI commands can provide access to processor registers and
memories, as well as controls such as soft reset, start, and stop.
GarageWorks also provides a shared configuration system that keeps design
parameters and AXI address mappings consistent across the DUT, the Chisel-based
AXI dispatcher, and the cocotb testbench. The AXI dispatcher and supporting
test infrastructure are written in Chisel for conciseness and parameterization,
while the DUT may be written in either Chisel or Verilog.

\subsection{Test Configuration and Backend Switching}

On the software side, GarageWorks provides a common bridge abstraction over the
underlying simulation or FPGA driver. The current implementation includes
\texttt{cocotb\_bridge} for software RTL simulation and
\texttt{aved\_bridge} for FPGA execution through \texttt{pyaved}. A helper
module in the Chisel-based AXI dispatcher generates a shared JSON configuration
file containing DUT design parameters and the AXI address map. Each bridge
loads this JSON file and exposes common primitive operations, such as soft reset,
AXI read, AXI write, and expected-value checks, to the shared testbench.

GarageWorks provides a source-to-source
translator built with Python's \texttt{LibCST} library to convert cocotb
testbench code into the FPGA-targeted form. This avoids manual
modification of the original testbench and allows users to switch back to RTL
simulation when debugging capability is needed.

On the host side, the current FPGA implementation is based on the Alveo Versal
Example Design (AVED). We extended AVED with a C API and provide a Python
abstraction, \texttt{pyaved}, using \texttt{ctypes}. The
\texttt{aved\_bridge} uses this interface to issue memory-mapped commands and
retrieve results from the FPGA.

\subsection{AI-Assisted Test Component Generation}

Even with the shared abstraction, each DUT still requires design-specific
components such as AXI dispatcher logic, address mappings, bridge
configuration, testbench code, and build files. GarageWorks therefore also
provides AI-assisted generation of DUT-specific test components, including AXI
dispatchers, address maps, and Chisel/cocotb testbenches. A clearly defined and
tested abstraction architecture significantly improves generation quality by
constraining the task. In contrast, open-ended generation of complete FPGA test
environments is unreliable because LLMs may not capture platform-specific
behaviors, integration constraints, or implementation workarounds that are
often learned through hardware bring-up.

We applied this approach to a custom RISC-V core and an X-ray online data-processing design. Given the DUT RTL and the GarageWorks base architecture, the AI generated the AXI parameters and address maps, Chisel AXI dispatcher logic, cocotb testbench components, and build files. The generated components worked without modification in both software RTL simulation and FPGA execution, and also serve as templates for developing more complex tests.

\section{Discussion}

GarageWorks' well-defined bridge abstraction improved productivity even with manual integration and also simplified AI-assisted generation.
The same testbench can be switched between RTL simulation and FPGA execution without manual modification.
We have also successfully tested AI-assisted generation of DUT-specific components for designs including a RISC-V core and an X-ray online data-processing design.
The open-source release of GarageWorks is available at \url{https://github.com/hwspec/garageworks}, including the AXI BFM, Chisel AXI dispatcher, parameter/address-map sharing mechanism, Python bridges, and examples.
The current implementation has been tested on AMD Alveo V80 FPGAs; we plan to support additional FPGA platforms and integrate formal verification, PPA estimation, and FPGA-side printf-like debugging.

\section*{Acknowledgments}

This work was supported by the U.S. Department of Energy, Office of Science,
under Contract No.~DE-AC02-06CH11357, and in part by the Office of Advanced
Scientific Computing Research Computer Science Competitive Portfolios program
under Contract No.~DE-AC02-05CH11231. The author thanks Argonne colleagues
Kyle Schmitt, Ben Allen, and Connor Bohannon for FPGA bring-up and
customization of the software stack, and Mike Papka and Valerie Taylor for
their support of this FPGA work.
Claude (Anthropic) was used for writing assistance and editing. The author reviewed and takes full responsibility for all content.

\newpage

\bibliographystyle{IEEEtran}
\bibliography{ref}

\end{document}